\documentclass{ws-procs10x7square}

\usepackage{amssymb,cite}
\newcommand{\PT}{{\mathcal P}{\mathcal T}}

\newcommand{\nlinsys}[1]{{\tt nlin-si/#1}}
\begin{document}


\title{Differential equations and the Bethe ansatz}

\author{Patrick Dorey}
\address{Department of Mathematical Sciences,\\
         University of Durham,\\
         Durham DH1 3LE,
         UK.\\
         p.e.dorey@durham.ac.uk}

\author{Clare Dunning}
\address{Department of Mathematics,\\
         University of York,\\
         York YO1 5DD,
         UK.\\
         tcd1@york.ac.uk}

\author{Adam Millican-Slater}
\address{Department of Mathematical Sciences,\\
         University of Durham,\\
         Durham DH1 3LE,
         UK.\\
         adam.millican-slater@durham.ac.uk}

\author{Roberto Tateo}
\address{Department of Mathematical Sciences,\\
         University of Durham,\\
         Durham DH1 3LE,
         UK.\footnote{{\rm
        \uppercase{A}ddress after \uppercase{N}ovember 
        2003: \uppercase{D}ipartimento di
            \uppercase{F}isica \uppercase{T}eorica, 
         \uppercase{U}niversit\`a di \uppercase{T}orino,
         \uppercase{V}ia \uppercase{P}.~\uppercase{G}iuria 1, 10125
         \uppercase{T}orino, \uppercase{I}taly.}}  
\\
         roberto.tateo@durham.ac.uk}


\maketitle

\abstracts{
We review some surprising links which have been
discovered in the last few years between the theory of certain
ordinary differential equations, and particular integrable lattice
models and quantum field theories in two dimensions. An application of
this correspondence to a problem in non-Hermitian (PT-symmetric)
quantum mechanics is also discussed.\\[3pt]
Contribution to the proceedings of the International Congress of 
Mathematical Physics, Lisbon 2003; 
preprint DCPT-03/45,
{\tt hep-th/0309054}.
}


\section{Introduction}
In the last few years, detailed connections have begun to emerge
between two previously-separated areas of mathematical physics:
the spectral properties of ordinary differential equations such
as the Schr\"odinger equation, and the study of integrable lattice
models and integrable quantum field theories using techniques related
to the Bethe ansatz. This is sometimes given the (perhaps over-grand)
title of the `ODE/IM correspondence'. While the correspondence remains
at the level of a mathematical coincidence, albeit a detailed one, it
has already provided some important new insights into a variety of
problems.

On the `ODE' (ordinary differential equations) side, some of the key themes
are Hermitian and non-Hermitian
spectral problems, Schr\"odinger equations, the WKB method, analytic
continuation and $\PT$-symmetry. These have been developed by a
variety of authors; some important names and references for the story that
we want to tell are
Sibuya~\cite{Sha}, Voros~\cite{Voros}, Bessis and Zinn-Justin~\cite{BZJ}, and
Bender and Boettcher~\cite{BB,BBN}.

The `IM' (integrable models) aspect has a similarly long history.
The theories relevant for the first example of the correspondence to
be discovered \cite{DTa} are
the six-vertex model, and the quantum field theory of the massless
sine-Gordon model. The Bethe ansatz, TQ relations and fusion
hierarchies all have a r\^ole to play, and the works on these topics by
Baxter~\cite{Bax}, Kl\"umper and Pearce~\cite{KP}, and Bazhanov, Lukyanov
and Zamolodchikov~\cite{BLZ1,BLZ2,BLZ3} are particularly relevant.

In a short survey such as this, it is impossible to do justice to all
of this background. In the next two sections some of the basic
vocabulary will be sketched.   More detailed reviews have been
given in
\cite{DDTrev,DDTrevb}, and an extended version is currently in
preparation.  Subsequent developments following the initial
observation in \cite{DTa} can be found in
\cite{BLZa,Sa,FAiry,DTb,DTc,Sb,Sc,DDTa,Srev,Hikami,
DDTb,DDTc,BHK,BLT,LZ}.

\section{First prologue: the basics of integrable models}
\label{prol1}
Consider a rectangular two-dimensional lattice, of size $N\times M$. On
each link of the lattice, attach an arrow, or {\em spin}, $\sigma$,
pointing in one of the two possible directions along the link. An
assignment of a spin to each link $i$ of the lattice gives a {\em
configuration}\/ $\{\sigma_i\}$. If we impose the additional
constraint that the number of arrows pointing into each vertex is
equal to the number pointing out, then we are on the way to defining
the six-vertex model.  The main task is to calculate
averages over all possible configurations, weighted by relative
probabilities which can be found as follows. First, the local {\em
Boltzmann weights}\/ $W[a,b,c,d]$ must be specified, one number
for each possible set of values taken by the
spins on the four links connected to a single
site of the lattice. Then the relative probability of a given
configuration is simply the product of these local weights over all
sites.
A key quantity is now the sum of these relative probabilities over all
possible configurations, which gives the normalisation factor for the
calculation of the actual probability of any quantity. Otherwise known
as the {\em partition function}, it is given schematically as
\begin{equation}
Z=\sum_{\{\sigma_i\}}~\prod_{\rm sites}W[a,b,c,d]~.
\end{equation}
The local Boltzmann weights $W$, and hence also the partition function
$Z$, may depend on physical parameters such as the temperature.
For the six-vertex model there is just one physically-relevant
quantity to specify, namely the anisotropy $\eta$. In the related
quantum field theories this becomes the sine-Gordon coupling $\beta$,
related to $\eta$ by $\beta^2=1-2\eta/\pi$.
For the solution technique we wish to describe below, however, a
rather less physical parameter will be especially important, for
reasons which will be sketched shortly: the {\em
spectral parameter}\,~$\xi$.

To evaluate $Z$, we can first define the
{\em transfer matrix}\/ $\mathbb T$, to be the sum over the spins on
a line of horizontal links of the product of their Boltzmann weights,
for a given set of spins on the adjacent vertical links above and
below. If we assign one multiindex $\alpha$ to the vertical links
above the line in question, and another $\alpha'$ to the links below,
then $\mathbb T$ can be thought of as a $2^N\times 2^N$ matrix
$\mathbb T_{\alpha'}^{\alpha}$, and, for periodic boundary conditions,
the partition function $Z$ is
\begin{equation}
Z={\rm Trace}\left[{\mathbb T}^M\right]~.
\end{equation}
Clearly, if we can diagonalise $\mathbb T$, then we shall be able to
compute $Z$. Even better, since the main interest is in the so-called
{\em thermodynamic limit}\/ when both $N$ and $M$ tend to infinity, we
only need find the lowest-lying eigenvalues. Note that the
mathematical structures
found in this limit can also be obtained directly in the context of
the continuum quantum field theory of the massless sine-Gordon model
on a cylinder,
using the constructions of
\cite{BLZ1,BLZ2,BLZ3}.

Now comes the first moment where the concept of {\em integrability}
arises: the local Boltzmann weights for the six-vertex model are so
defined that the transfer matrices taken at different values of the
spectral parameter commute:
\begin{equation}
[{\mathbb T}(\xi),{\mathbb T}(\xi')]=0~.
\end{equation}
(The deeper meaning of this property would
take us too far afield, but from one point of view it is related to the
existence of infinitely-many
commuting conserved quantities.) If the matrices ${\mathbb T}(\xi)$
all commute, then they can be diagonalised simultaneously and we can
study individual eigenvalues $T(\xi)$ as functions of $\xi$. There are
various methods for finding these functions, but the most relevant
here was discovered by Baxter in the early 1970s, in connection with
his work on the more-complicated eight-vertex model. Since the
transfer matrix is an entire function of $\xi$ with $\xi$-independent
eigenvectors, the eigenvalues $T(\xi)$ must also be entire. Baxter
showed further that, for each eigenvalue $T(\xi)$, there is an
associated function $Q(\xi)$, also entire, such that the following
{\em TQ relation}\/ holds:
\begin{equation}
T(\xi)Q(\xi)=e^{-2\pi ip}Q(q^{-2}\xi)+e^{2\pi ip}Q(q^2\xi)
\label{tqeq}
\end{equation}
where the phase $q$ depends on  $\beta$
as $q=e^{i\pi\beta^2}$, and the extra parameter $p$ allows for the
possibility of inserting a `twist' into the periodic
boundary conditions on the
lattice.
(Strictly speaking, for finite values of the lattice width $N$
the TQ relation contains some extra factors, but these go in the
thermodynamic limit. We are following the
conventions of
\cite{BLZ2}, appropriate for this limit, save that
there 
$Q$ is denoted $A$ for $p\neq 0$, and both it
and $T$ are taken to be entire functions
of the {\em squares}\/ of their arguments.)
To see why this equation provides a useful
constraint, we can argue as follows. Since $Q$ is entire, it is,
subject to suitable growth conditions, determined by the location of
its zeroes. Suppose these zeroes are at $\{e_0\dots e_k\}$\,; then
\begin{equation}
Q(\xi)=Q(0)\prod_{i=0}^k\left(1-\frac{\xi}{e_i}\right)\,.
\label{prodform}
\end{equation}
Now $T(\xi)$ is also entire, so the TQ relation (\ref{tqeq}) taken at
$\xi=e_j$ implies that $e^{-2\pi ip}Q(q^{-2}e_j)+
e^{2\pi ip}Q(q^2e_j)=0$. Rearranging and
using the product form (\ref{prodform}),
\begin{equation}
\prod_{i=0}^k\left(\frac{e_i-q^2e_j}{e_i-q^{-2}e_j}\right) = -e^{-4\pi
i p}\,.
\label{bae}
\end{equation}
This gives us $k{+}1$ equations in $k{+}1$ unknowns, which determine
the $e_i$ up to discrete ambiguities corresponding to the fact that
the transfer matrix has not one but many eigenvalues. These equations
coincide with the so-called {\em Bethe ansatz equations}\/ (BAE) found in
other, more direct, approaches to the problem. In this context, the
$e_i$ are sometimes called the {\em Bethe roots}. Note, once the Bethe
roots are
known, the eigenvalue $T(\xi)$ is easily recovered by using
(\ref{prodform}) and then (\ref{tqeq}). Although the BAE (\ref{bae}) do not
determine the Bethe roots uniquely, for the so-called ground state
eigenvalue -- the largest one, most important in the thermodynamic
limit -- this ambiguity is fixed by the fact that the Bethe roots for
this eigenvalue are all located on the positive real axis.

Over the last few decades, Bethe ansatz equations such as (\ref{bae})
have been studied extensively, and
much is known about their solutions.  For the moment the key thing to
remember is the simple way that they follow from the TQ relation. If
a similar functional equation can be found in another context, we can
hope to establish a rather detailed connection with the
already-existing body of knowledge on integrable models.

\section{Second prologue: ${\PT}\!$-symmetric quantum mechanics}
We now make a change of tack, and turn our attention to some
quantum-mechanical problems first investigated by Bessis and
Zinn-Justin in the early 1990s \cite{BZJ}, and then successively
generalised in \cite{BB,BBN,DTb,DDTb,DDTc}.
For the current discussion it will suffice to consider the following
collection of non-Hermitian `position space'
Hamiltonians:
\begin{equation}
{\mathcal H}_{M,l}=-\frac{d^2}{dx^2}+(ix)^{2M}+\frac{l(l{+}1)}{x^2}~,
\label{geneq}
\end{equation}
where $M$ is a positive real number.
The interest is in the spectrum of ${\mathcal H}_{M,l}$\,:
those values of $E$ such that the ordinary differential equation
${\mathcal H}_{M,l}\psi=E\psi$ has a normalisable solution. By
`normalisable', for $M<2$ we simply mean square-integrable on a
contour running along the real
axis, distorted just below the origin to avoid the singularity there
when $l(l{+}1)\neq 0$.
For $M\ge 2$, the contour must be further distorted to ensure
the correct analytic
continuation, as explained in~\cite{BB}.

\begin{figure}[ht]
\[
\begin{array}{cc}
{\epsfxsize=.37\linewidth%
\epsfbox{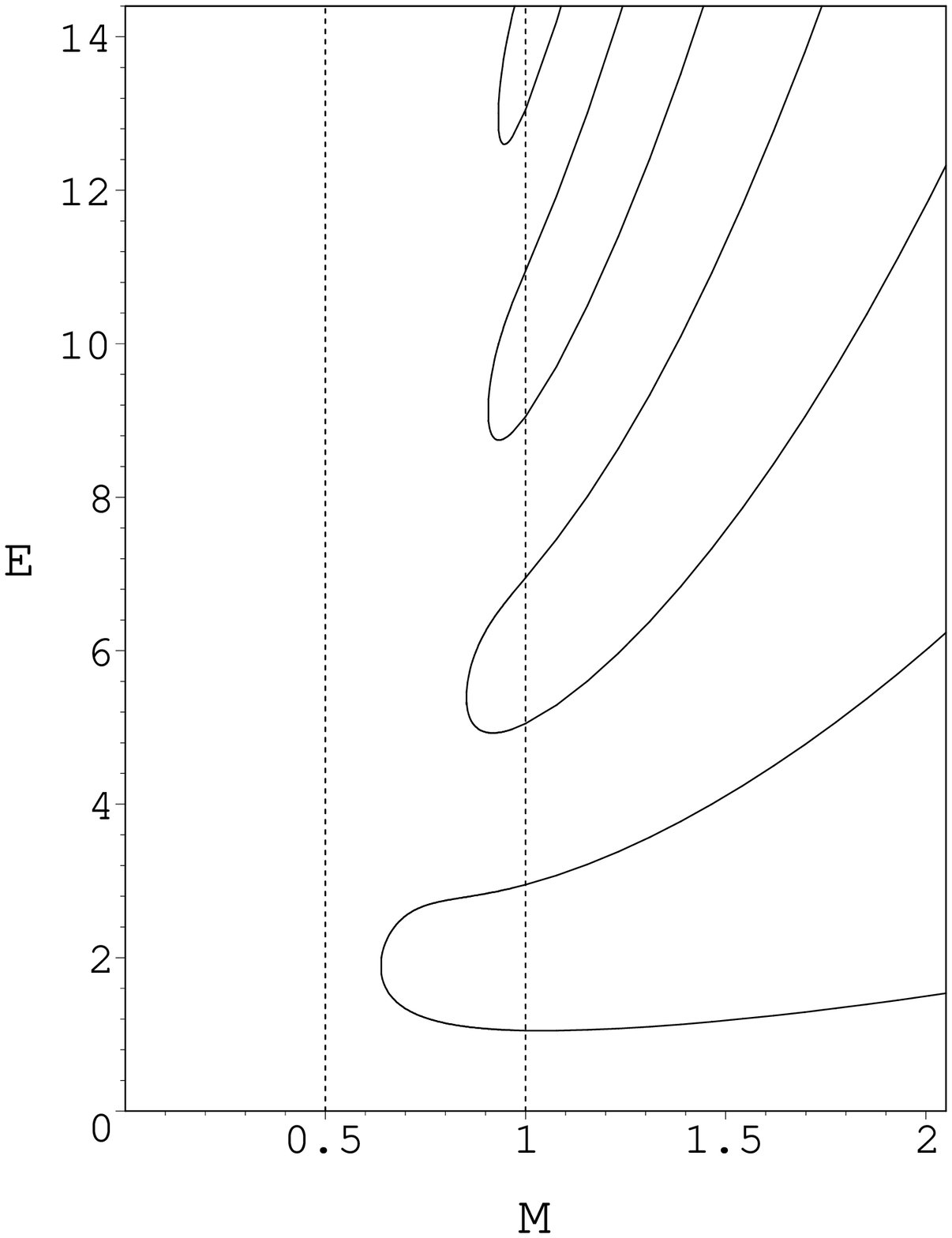}}~~~~~~~&~~~~~~~
{\epsfxsize=.37\linewidth%
\epsfbox{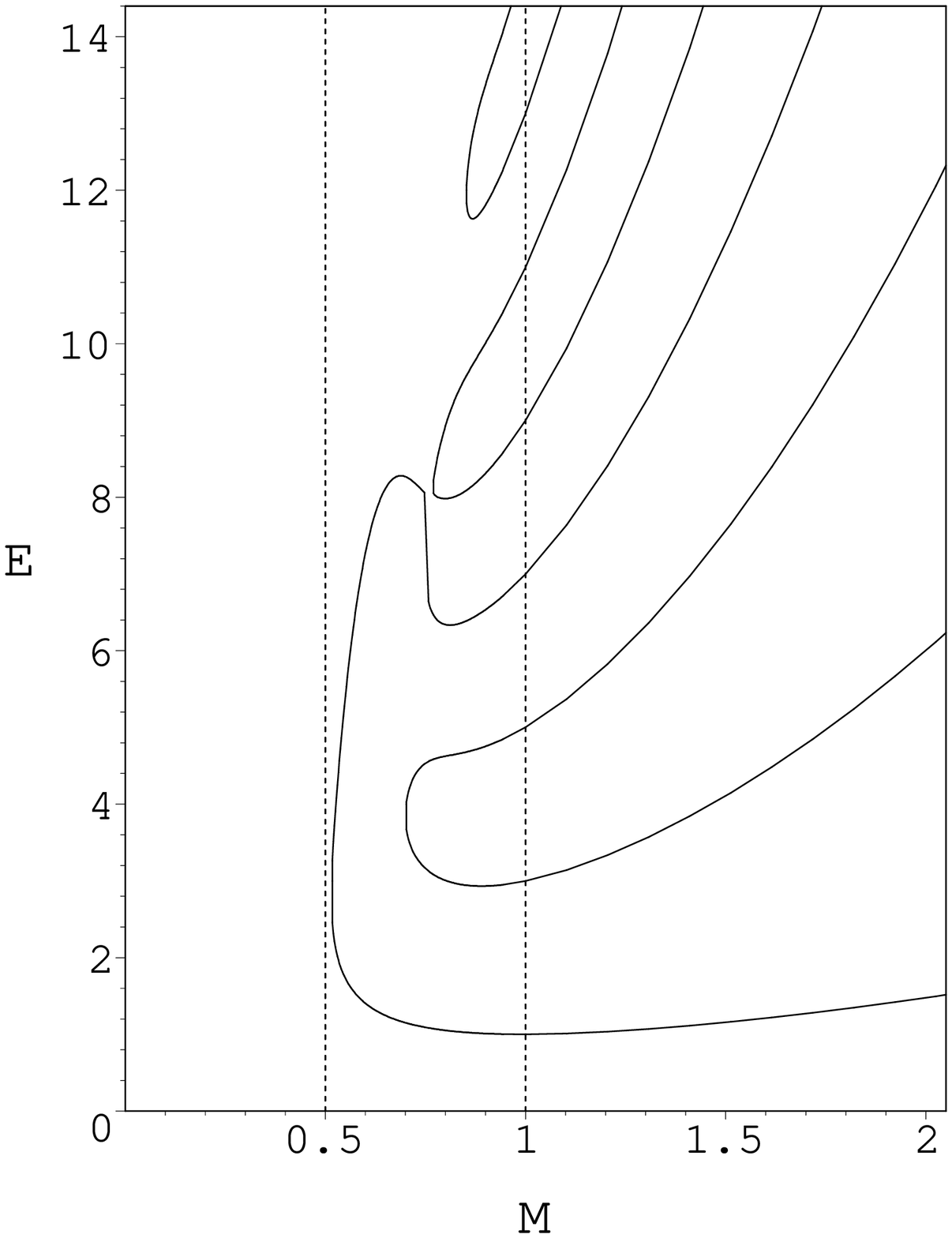}}\\[0pt]
\parbox{0.4\linewidth}{~~~~~\protect\footnotesize\ref{numerics1}a:
$l=-0.025$}&
\parbox{0.4\linewidth}{~~~~~~~~~~~~~\protect\footnotesize\ref{numerics1}b:
$l=-0.001$}
\end{array}
\]
\caption{Eigenvalues of ${\mathcal H}_{M,l} \psi = E\psi$, plotted as
a function of $M$ at $l=-0.025$ and $l=-0.001$.}
\label{numerics1}
\end{figure}

In figure \ref{numerics1} we show the spectrum as a function of $M$
for two negative values of $l$. Note that $M=1$ corresponds to the
exactly-soluble simple harmonic oscillator with angular momentum, and
that this point marks a profound change in the nature of the
spectrum: for $M\ge 1$ it is entirely real, while for $M<1$
infinitely-many eigenvalues pair off and become complex. The reality
of the spectrum for $M=3/2$, $l=0$ was the subject of Bessis and
Zinn-Justin's original conjecture \cite{BZJ}, which was extended to all
$M\ge 1$, $l=0$ by Bender and Boettcher in \cite{BB}. However a
complete proof was surprisingly elusive, and has only recently been
given, making essential use of the ODE/IM correspondence \cite{DDTb}.

\begin{figure}[ht]
\[
\begin{array}{cc}
{\epsfxsize=.37\linewidth%
\epsfbox{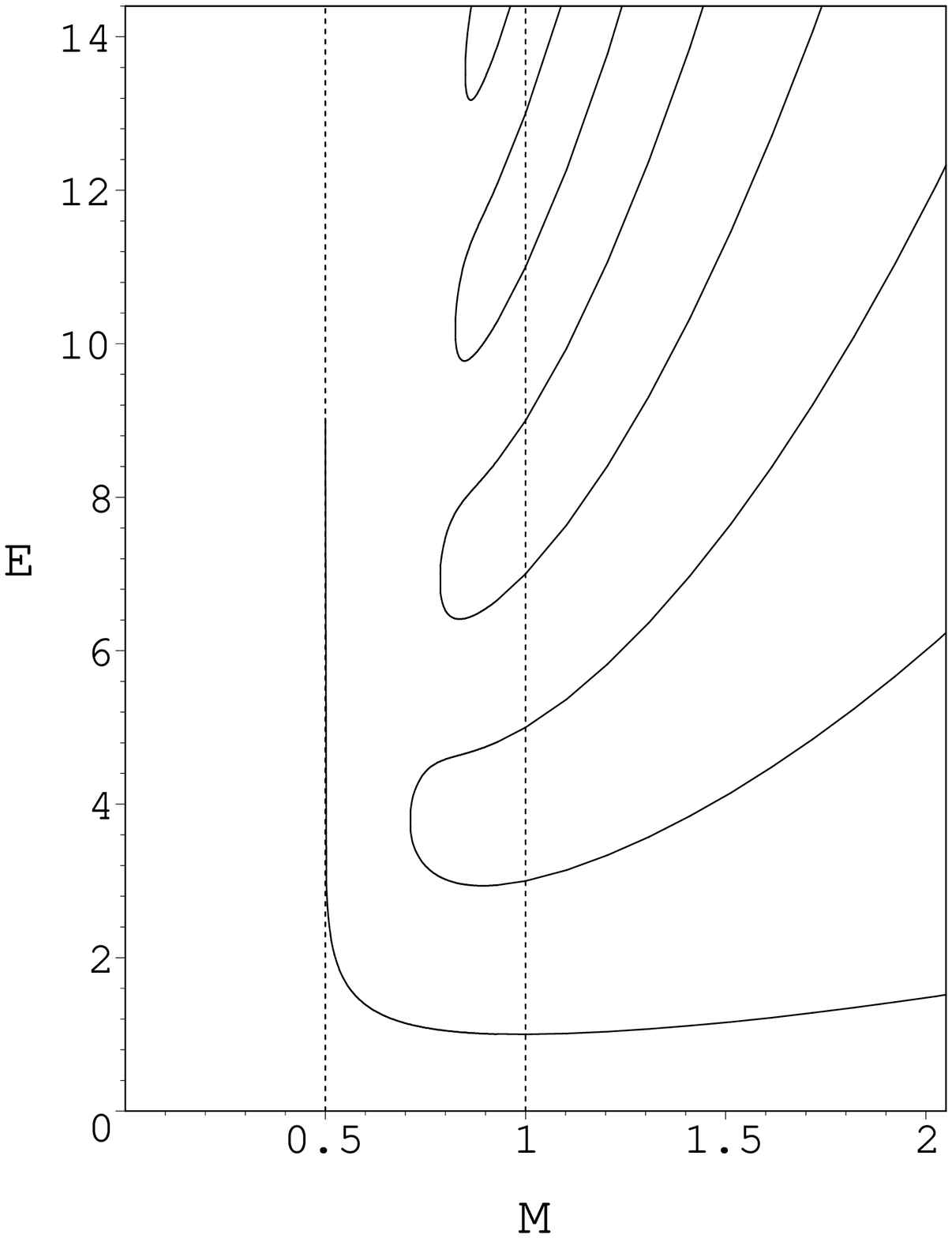}}~~~~~~~&~~~~~~~
{\epsfxsize=.37\linewidth%
\epsfbox{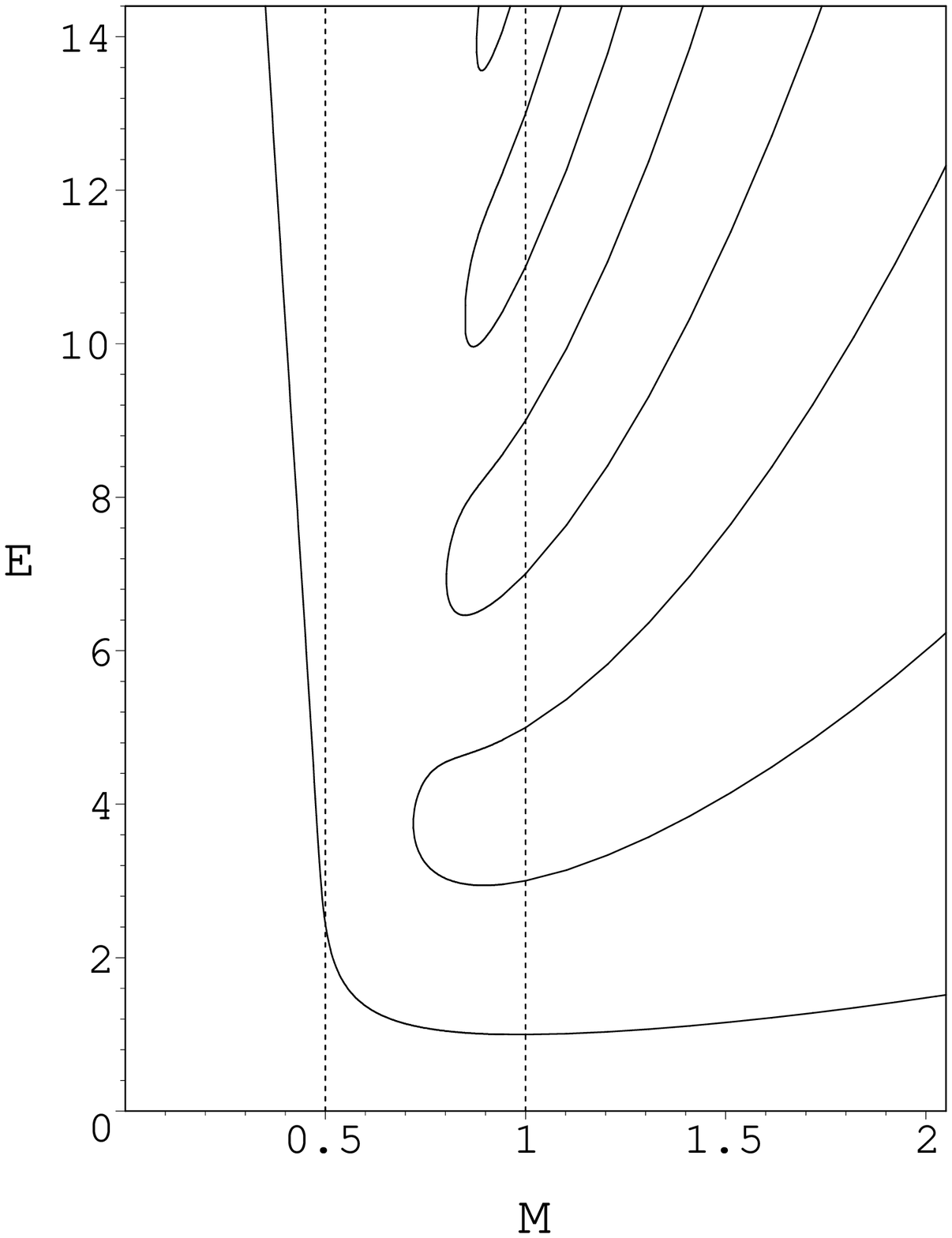}}\\[0pt]
\parbox{0.4\linewidth}{~~~~~\protect\footnotesize\ref{numerics2}a:
$l=0$}&
\parbox{0.4\linewidth}{~~~~~~~~~~~~~\protect\footnotesize\ref{numerics2}b:
$l=0.001$}
\end{array}
\]
\caption{Eigenvalues of ${\mathcal H}_{M,l} \psi = E\psi$, plotted as
a function of $M$ at $l=0$ and $l=0.001$.}
\label{numerics2}
\end{figure}

The nature of the transition to complex eigenvalues
as $M$ falls below $1$ looks rather simple
on figure \ref{numerics1}a, but less so once figure \ref{numerics1}b
has been
examined. In fact, as shown in figure \ref{numerics2}, for $l\ge 0$
the connectivity pattern seen in figure \ref{numerics1}a is completely
reversed. Some further discussion of this behaviour can be found in
\cite{DTb,DMT}, while the even-richer structure which emerges when a
suitably-chosen inhomogeneous term is added to the potential is
discussed in \cite{DDTb,DDTc}.

\vspace{-0.05cm}

As emphasised in \cite{BB,BBN}, a key feature of the
Hamiltonians (\ref{geneq}) is that, while not Hermitian, they are
unchanged by the combined action of parity ${\mathcal P}$ and
time-reversal ${\mathcal T}$ -- whence the name $\PT$-symmetric
quantum mechanics. 
We hope to have
demonstrated that the
subject contains many interesting phenomena for the mathematical
physicist to explore;
see
\cite{DP,Za,Mez,Zl,DT,BCQ,bender:wkb,BW,Mez1,
HX,TF,BBJMS,CIRR,GS,KS,AM,Bender:2002vv, Mosta:cpt, BMWhid,weig2} and
references therein for some further developments in the area.
However, from the point of view of this article,
its main interest lies in its surprising links with the six-vertex
model discussed in section \ref{prol1} above, which we shall now
describe.

\section{Spectral determinants and the correspondence}

Rather than worrying about individual eigenvalues of (\ref{geneq}),
it turns out to be useful to combine them into a single function -- a
spectral determinant -- and then exploit the analytic properties
of this function. This approach was particularly advocated in the
differential equation context by Sibuya \cite{Sha} and Voros
\cite{Voros}. Here, influenced by parallels with the theory of
integrable models, we shall use spectral determinants to
couple the non-Hermitian problem
discussed so far with a related Hermitian problem, obtained
from
(\ref{geneq}) by sending $x \to x/i$ and $E \to -E$. The new
 problem is
\begin{equation}
\left(-\frac{d^2}{dx^2}+x^{2M}+\frac{l(l{+}1)}{x^2}\right)\Phi(x) = E
\Phi(x)~,
\label{eqd}
\end{equation}
and we choose as boundary conditions
that the solution should vanish as $x \to \infty$ along the real axis, and
behave as $x^{l+1}$ as $x \to 0$.   For $\Re e\,l>-1/2$, this problem is
Hermitian. In the language of ordinary differential equations in the
complex domain, it is sometimes called a `radial' problem, while
(\ref{geneq}) is called a `lateral' problem.
(Note, the radial problem can also be considered for $\Re e\,l\le
-1/2$,
but is best then defined by analytic continuation in $l$.)

Let $\{E_j\}$ be the set of eigenvalues of
(\ref{geneq}), and let $\{e_i\}$ be the
eigenvalues of (\ref{eqd}). Then define
two spectral determinants, as follows:
\begin{equation}
T(E) =T(0)\prod_{j=0}^{\infty}\left(1+\frac{E}{E_j}\right)~,
\label{specdetT}
\end{equation}
and
\begin{equation}
Q(E) =Q(0)\prod_{i=0}^{\infty}\left(1-\frac{E}{e_i}\right)~.
\label{specdeta}
\end{equation}
Both products are convergent for $M>1$, and define
entire functions of $E$, and
their zeroes (or, for $T(E)$, their negatives)
coincide with the eigenvalues of the
corresponding spectral problem. For $M\le 1$
convergence factors must be added, and it is more efficient
to define the spectral determinants indirectly, via certain special
`Sibuya' solutions to (\ref{eqd}), which are anyway needed to prove
the key identity (\ref{tqeqb}) below; see \cite{Sha,DTb} for more details.

By considering the asymptotic behaviour of the Sibuya solutions in
the complex plane it is
not hard, following the arguments given in \cite{DTb},
to establish a Stokes relation, from which
one can obtain the following functional equation
\begin{equation}
T(E)Q(E)= \omega^{-(2l+1)/2} Q(\omega^{-2}E) +
 \omega^{(2l+1)/2} Q(\omega^{2}E)~,
\label{tqeqb}
\end{equation}
where $\omega= e^{i\pi/(M+1)}$. This shows that the spectral problems
(\ref{geneq}) and (\ref{eqd}) are related by much more than a simple
change of variables and boundary conditions; and furthermore, that
this relation is encoded precisely in a TQ relation of the sort which
had previously arisen in the context of the six-vertex and sine-Gordon
models.

\section{Consequences of the correspondence}

If we set $\beta^2=1/(M{+}1)$ and $p=(2l{+}1)/(4M{+}4)$,
then the two TQ relations (\ref{tqeq}) and (\ref{tqeqb})
match perfectly.
The remaining ambiguity in the solutions of the resulting
Bethe ansatz equations is resolved once we recall that
the problem (\ref{eqd}) is Hermitian, and so all of its eigenvalues
$e_i$ are real. Given the remarks about the locations of Bethe roots
at the end of section 2, this
means that the spectral determinant $Q(E)$ arising in the ordinary
differential equation should be identified with the
Q-function $Q(\xi)$ for the ground state of the sine-Gordon model.
(We must also match the normalisations of $E$ and $\xi$, since these
are left undetermined by the TQ relations.
This is easily achieved by comparing the leading
asymptotic of the field theory
Q-function given in \cite{BLZ2} with that obtained
by a simple WKB analysis of the ordinary differential equation -- see
\cite{DTa,DTb} for more details.) Once the Q-functions have been
identified, so can be the T-functions, and we have our main punchline:

\begin{quote}
The spectral determinant $T$\/ for the lateral
($\PT$-symmetric)
problem (\ref{geneq}) is equal to the ground state, or
vacuum, eigenvalue of the
transfer matrix $\mathbb T$ of the massless sine-Gordon model on a
cylinder.
\end{quote}

\noindent
Together with the corresponding statement for $Q$, this provides
a powerful tool to analyse the functions constructed in
the context of integrable quantum field theory in
\cite{BLZ1,BLZ2,BLZ3}, as the analytic properties of solutions of
ordinary differential equations are relatively easy to control.
(In this regard, we should mention that, for a complete proof of the
correspondence, it currently seems best to argue via the so-called
`quantum Wronskian' relation, as in \cite{DTa,BLZa}. However the route
we have adopted here is perhaps more intuitive, and it leads naturally on
to the reality proof which we shall describe shortly.)
There are also less direct applications of the correspondence -- 
for example, it
yields a proof \cite{BLZa} of a duality relation in quantum Brownian
motion, first proposed in \cite{FZ}. In finishing this contribution,
we mention two more, both on the ordinary
differential equations side of the correspondence.

The first concerns the reality properties of the $\PT$-symmetric
problems (\ref{geneq}). We have already seen, in section 2, the
argument which shows that
substituting $E=e_j$ into the TQ relation (\ref{tqeqb})
leads to Bethe ansatz
equations, which in the light of the correspondence we can
reinterpret as coupling the different eigenvalues of the Hermitian
problem (\ref{eqd}). However, if we instead set $E=-E_j$, the same
reasoning leads, in the region $M>1$ for which the factorised form
(\ref{specdeta})
applies, to
\begin{equation}
\prod_{i=0}^{\infty}\left(\frac{e_i+\omega^2E_j}{e_i+\omega^{-2}E_j}\right)
 = -\omega^{-2l-1}\,.
\label{bae2}
\end{equation}
This equation couples the so-far mysterious eigenvalue
$E_j$
of the $\PT$-symmetric problem (\ref{geneq}) to the much
better-controlled, and indeed real and positive, eigenvalues $e_i$ of
the Hermitian problem (\ref{eqd}). If we take the modulus$^2$ of
both sides, it is soon seen that the only way that the resulting
equality can be
achieved is for $E_j$ to be real, as had previously been conjectured.
Notice, the proof
breaks down for $M\le 1$, as has to be the case given the numerical
findings of
\cite{BB,DTb}, illustrated in figures \ref{numerics1} and
\ref{numerics2} above.
More details, in a slightly more general setting than
that described here, can be found in \cite{DDTb}; for a further
generalisation of the method, see \cite{Shin:2002vu}.

The second application was also found in \cite{DDTb}, but is
a little more specialised. We consider the
radial problem at $M=3$, and add an inhomogeneous term $\alpha x^2$
to the potential. Although this was not done in the
original paper, it will also be convenient to reparametrise the
angular-momentum term by setting $\lambda=\sqrt{3}(2l{+}1)$\,, so that
the equation becomes
\begin{equation}
\left(-\frac{d^2}{dx^2}+x^{6}+\alpha x^2+
\frac{\lambda^2-3}{12\,x^2}\right)\Phi(x) = E
\Phi(x)~.
\label{eqe}
\end{equation}
Via the Bethe ansatz approach, it turns out that this problem has a
relationship with a {\it third}-order ordinary differential
equation:
\begin{equation}
\left( D(g_2{-}2)D(g_1{-}1)D(g_0) + x^3\right)\phi=
\frac{3\sqrt{3}}{4}\,E\phi
\end{equation}
where $D(g)\equiv (\frac{d}{dx}-\frac{g}{x})$, and
\begin{equation}
g_0=1+(\alpha+\sqrt{3}\lambda)/4~~,~~
g_1=1+\alpha/2~~,~~
g_2=1+(\alpha-\sqrt{3}\lambda)/4~.
\end{equation}
This third-order equation is associated with $SU(3)$ Bethe ansatz
equations, as discussed in \cite{DTc,DDTa}. Furthermore, the
third-order equation is symmetrical in $\{g_0,g_1,g_2\}$, a
feature which is completely hidden in the original second-order
equation. By playing with this symmetry, one can establish some
novel spectral equivalences between different
(second-order) radial problems, and also between these and certain
lateral problems. More details and explicit formulae can be found
in \cite{DDTb}, and here we simply add to that paper
the remark that, when expressed
in terms of the variables $(\alpha,\lambda)$, the mappings turn out to
act as certain $2\times 2$ matrices in the Weyl group of $SU(3)$.

Other aspects of the ODE/IM correspondence are still being developed
as this contribution is being written. It would be nice to have a
working correspondence also for finite lattice systems, before the
thermodynamic limit is taken, and some small progress in this
direction will be reported in \cite{DST}.
Another important question that until recently remained open
concerns the
other states in the quantum field theories,
besides the ground state, and whether they
can also be matched with differential equations.
In a very recent paper \cite{BLZhigh}, Bazhanov, Lukyanov and
Zamolodchikov have answered this in the affirmative. The
`Q-potentials' that they construct are no longer real, even for the
radial problems. This is not as surprising as it might seem --
recall that it is
only for the ground state that the Bethe roots, which correspond to
the eigenvalues of the radial spectral problem, all lie on the real axis.
For other states there are complex roots, and so even the radial
problems can no longer be Hermitian.

Finally, it would
be valuable to have a more physical understanding of the relationship
between integrable
quantum field theories and ordinary differential equations.
At the moment this remains
mysterious, but sufficiently-many examples of the phenomenon have now been
collected that progress may not be too far off.


\section*{Acknowledgments}

PED thanks the organisers for the invitation to speak at the
conference; TCD, AM-S and RT thank the UK EPSRC for a Research
Fellowship, a Studentship
and an Advanced Fellowship respectively.
This work was partially supported by the EC network ``EUCLID'',
contract number HPRN-CT-2002-00325.






\end{document}